\documentclass[10pt,letterpaper]{article}
\usepackage[utf8]{inputenc}
\usepackage{amsmath}
\usepackage{amsfonts}
\usepackage{amssymb}
\usepackage{multirow}
\usepackage{bm}
\usepackage{xcolor}
\usepackage[english]{babel}
\usepackage{graphicx}
\usepackage{hyperref,cite}
\hypersetup{
    colorlinks=true,
    linkcolor=blue,
    filecolor=blue,  
    citecolor=blue,
    urlcolor=blue,
    }
\usepackage{lipsum}

\title{Unified Lorentz-covariant Poisson Bracket for\\ the Electrodynamics of a Point Particle}

\author{J. F. Pérez-Barragán\footnote{Contact email address: jfcoperezb@estudiantes.fisica.unam.mx}\\
\normalsize{Instituto de Física, Universidad Nacional Autónoma de México, 04510 Mexico City, Mexico}}

\date{}

\begin{document}

\maketitle

\abstract{\noindent Using the multisymplectic Hamiltonian formalism, we propose a Poisson bracket for an electromagnetic radiation field that, in addition to satisfying the restricted principle of relativity, reproduces well-established results from the conventional formulations of field theories. To this end, it is necessary to translate the multisymplectic description of the system to the field's momentum representation, where a bilinear and Lorentz-covariant form emerges in a straightforward manner. Consequently, we establish a description for a system consisting of an electromagnetic radiation field and a point particle interacting with each other that is based on a single Lorentz-covariant Poisson bracket with respect to the canonical variables of the system. The results obtained from this proposal offer a solid foundation for the development of a Lorentz-covariant quantization process.

\vspace{0.25cm}
}

\section{Introduction}\label{Sec1}

Since the early objections raised by Born \cite{MB1934}, the incompatibility between the requirement of full consistency with the restricted principle of relativity and the use of canonical quantization in modern field theories has been addressed on numerous occasions without reaching yet a definitive solution. The crux of the matter lies in the fact that canonical quantization requires a Hamiltonian (or canonical) formulation of field theories and therefore implies a primacy of time as fundamental independent variable \cite{PD1927,WH1929}. This incompatibility is not fundamental: quantum field theories constructed through canonical quantization do describe relativistic phenomena; however, their covariance must be verified at every stage of construction, typically through laborious proofs.

Alternative quantization processes that avoid privileging time and thus yield quantum field theories in a manifestly Lorentz-covariant manner have been developed over the years. The functional (or path-integral) quantization process, for example, formulates quantum field theories in terms of integrals over field configurations using the action of the system as weight function \cite{RF1948,JS1951,JS1953}. Unlike the Hamiltonian function, the action is a Lorentz scalar, so functional quantization preserves covariance throughout the formulation of the theories. Moreover, the Faddeev–Popov procedure, Becchi–Rouet–Stora–Tyutin quantization, and the Batalin–Vilkovisky formalism provide systematic, manifestly Lorentz-covariant methods for quantizing gauge theories \cite{LF1967,CB1976,IB1981,IB1983}. The Haag–Kastler axiomatic framework, in which fields are defined as operator-valued distributions subject to conditions that include covariance, offers, at least in principle, another Lorentz-covariant route for formulating quantum field theories \cite{RH1964}. Yet, despite these alternatives, constructing a fully Lorentz-covariant canonical quantization process remains an open problem in theoretical physics.

The starting point for addressing this problem has historically been the development of a Lorentz-covariant generalization of the Hamiltonian formalism. In this regard, a strong candidate for such a generalization is the multisymplectic Hamiltonian formalism (MHF), whose main feature is to manifestly preserve the covariance of the Hamiltonian description of fields through the introduction of a generalized momentum per field derivative. The seminal papers on the MHF were published by de Donder \cite{Donder1935} and by Weyl \cite{HW1935} in the 1930s. Nonetheless, after nearly a century of existence, the use of this formalism to construct classical or quantum field theories is not common practice, since its development beyond the basic definitions still presents formal lacunae \cite{MF2002}. In particular, an important drawback of the MHF is that it allows to define more than one Poisson bracket (PB) \cite{MF2002,JM1986,JC1991,MM1998,MM2001,FH2004a,MF2005,VG2006,JS2008,MF2015,LR2020}, which significantly dissipated the initial enthusiasm for using this formalism as the basis for developing a Lorentz-covariant canonical quantization.

Nevertheless, it has recently come to our attention that a Lorentz-covariant PB can be constructed in a straightforward manner for the electromagnetic radiation field using the MHF, by translating the multisymplectic description of the field into its momentum representation \cite{JFPB2025}. Additionally, and in a completely independent line of research, recent studies by Cetto et al. have established, within the framework of stochastic electrodynamics, a connection between the matrix formalism of quantum mechanics and the response functions of a particle interacting with an electromagnetic radiation field containing the zero-point radiation field, the non-thermal and stochastic electromagnetic background field that exists even in the absence of matter (for a review of stochastic electrodynamics up to 2015, see Ref. \cite{TheEmergingQuantum}). This connection is demonstrated by expressing the canonical variables of the mechanical subsystem in terms of the normal variables of the field’s modes of oscillation using PB. Thus, in the stationary phase of the interaction, where the field fully governs the particle’s motion, the kinematic transformation that changes the meaning of the quantities involved in the description of the particle becomes manifest \cite{Cetto2021,Cetto2022a,Cetto2022b}. Furthermore, the creation and annihilation operators of the radiation field have been deduced in similar manner, thereby providing a physical foundation for canonical quantization \cite{Cetto2023}.

These results motivate us to revisit the problem of constructing a Lorentz-covariant canonical quantization process using the MHF, since a relativistic generalization of Cetto's approach to quantization applicable to quantum field theories, most notably quantum electrodynamics, would not only physically justify the necessity for using operators instead of functions to describe fields, but also shed light on how to achieve a Lorentz-covariant canonical quantization. Hence, the purpose of the present work is to explore the possibility of describing a system composed of an electromagnetic radiation field and a point particle by means of a single Lorentz-covariant PB as a first step for developing the aforementioned generalization.

The paper is organized as follows. In Sect. \ref{Sec2}, after briefly reviewing the fundamentals of describing a system composed of an electromagnetic radiation field and a single point particle using the MHF, we translate the generated description to the momentum representation of the field. The central part of the paper is contained in Sect. \ref{Sec3} and deals with the proposal of a PB with respect to the canonical variables of the field that is capable not only of satisfying the restricted principle of relativity but also of reproducing well-established results from the conventional formulations of field theories. Subsequently, in Sect. \ref{Sec4}, the non-covariant form of the proposal is analyzed and compared with the standard (non-covariant) PB that is associated with an electromagnetic radiation field. Lastly, we outline a description of the complete system on the basis of the found PB and briefly discuss its main implications in Sect. \ref{Sec5}.

\section{Multisymplectic Hamiltonian Description}\label{Sec2}

We start by properly setting up the system to be investigated as an electromagnetic radiation field and a single point particle interacting with each other in a four-dimensional Minkowski space-time, with metric tensor $\eta_{\mu\nu}$ and signature
$(+,-,-,-)$. The action of such a system can be stated as
\begin{equation}
    S=\int{d}^{4}x\hspace{0.5mm}(\mathcal{L}_{\mathrm{field}}+\mathcal{L}_{\mathrm{part}}+\mathcal{L}_{\mathrm{int}}),\label{action}
\end{equation}
where $\mathcal{L}_{\mathrm{field}}$, $\mathcal{L}_{\mathrm{part}}$, and $\mathcal{L}_{\mathrm{int}}$ are the Lagrangian densities associated with the
free propagation of the field, the free motion of the particle, and the particle$+$field interaction, respectively. Thus, we establish that
\begin{equation}
    \mathcal{L}_{\mathrm{field}}=-\frac{1}{16\pi{c}}(F_{\mu\nu}F^{\mu\nu}+2\zeta(\partial_{\mu}A^{\mu})^{2}),\label{LagrangianField}
\end{equation}
\begin{equation}
    \mathcal{L}_{\mathrm{part}}=-\frac{m_{0}}{2}\int{d}\tau\hspace{0.5mm}\dot{u}_{\mu}\dot{u}^{\mu}\delta^{(4)}(x_{\alpha}-u_{\alpha}),
\end{equation}
and
\begin{equation}
    \mathcal{L}_{\mathrm{int}}=-\frac{e}{c}A_{\mu}\int{d}\tau\hspace{0.5mm}\dot{u}^{\mu}\delta^{(4)}(x_{\alpha}-u_{\alpha}),
\end{equation}
with $\zeta$ a positive, real parameter, $A_{\mu}$ the electromagnetic vector potential, $F_{\mu\nu}$  the electromagnetic tensor, and $m_{0}$, $e$, $u_{\mu}$, and $\tau$ the rest mass, electric charge, position, and proper time of the point particle, respectively. As usual, the movement of the particle is subject to the constraint $\dot{u}_{\mu}\dot{u}^{\mu}=c^{2}$. We have included a gauge-fixing term in the Lagrangian density $\mathcal{L}_{\mathrm{field}}$ in order to illustrate the descriptive capacity of the MHF

A canonical and manifestly Lorentz-covariant description of the system can then be achieved by associating the generalized (multisymplectic) momentum
\begin{equation}
    \theta_{\mu\nu}=-\frac{1}{4\pi{c}}(F_{\mu\nu}+\zeta\eta_{\mu\nu}\partial_{\alpha}A^{\alpha})\label{DefMomEM}
\end{equation}
to the electromagnetic vector potential and the well-known canonical momentum
\begin{equation}
    p_{\mu}=-m_{0}\dot{u}_{\mu}-\frac{e}{c}A_{\mu}(u_{\alpha})
\end{equation}
to the position of the particle, so that the action of the system can be written as
\begin{equation}
    \int{d}^{4}x\hspace{0.5mm}\left(\theta_{\mu\nu}\partial^{\mu}A^{\nu}+\int{d}\tau\hspace{0.5mm}p_{\mu}\dot{u}^{\mu}\delta^{(4)}(x_{\alpha}-u_{\alpha})-\mathcal{H}\right),\label{action2}
\end{equation}
where
\begin{equation}
    \mathcal{H}=-\frac{1}{2}\pi{c}\hspace{0.25mm}\theta_{\mu\nu}\left[\left(1+\frac{1}{2\zeta}\right)\theta^{\mu\nu}-\left(1-\frac{1}{2\zeta}\right)\theta^{\nu\mu}\right]-\frac{1}{2m_{0}}\int{d}\tau\left[p_{\mu}+\frac{eA_{\mu}}{c}\right]\left[p^{\mu}+\frac{eA^{\mu}}{c}\right]\delta^{(4)}(x_{\alpha}-u_{\alpha}),\label{Hamil1}
\end{equation}
is the so-called de Donder–Weyl function, i.e., the Lorentz-covariant (multisymplectic) Hamiltonian density \cite{Donder1935,HW1935}. On this basis, the Lorentz-covariant Hamilton's equations
\begin{equation}
    \partial_{\mu}A_{\nu}=\frac{\partial\mathcal{H}}{\partial\theta^{\mu\nu}},\quad\partial^{\mu}\theta_{\mu\nu}=-\frac{\partial\mathcal{H}}{\partial{A}^{\nu}},
\end{equation}
and
\begin{equation}
    \int{d}\tau\hspace{0.5mm}\dot{u}_{\mu}=\frac{\partial}{\partial{p}^{\mu}}\int{d}^{4}x\hspace{0.5mm}\mathcal{H},\quad\int{d}\tau\hspace{0.5mm}\dot{p}_{\mu}=-\frac{\partial}{\partial{u}^{\mu}}\int{d}^{4}x\hspace{0.5mm}\mathcal{H},\label{EcsHamiltonxpparticle}
\end{equation}
consistently reproduce the manifestly Lorentz-covariant forms of Maxwell's equations and the equation of motion of the particle, respectively.

Now, another possible Lorentz-covariant description of the system can be constructed in terms of the normal modes of oscillation of the field by expressing the retarded solution of Maxwell's equations as
\begin{equation}
    A_{\mu}=\frac{1}{8\pi^{3}}\lim_{m\rightarrow0}\left[\delta_{\mu}^{\nu}+\frac{\zeta-1}{\zeta}\frac{d}{dm^{2}}\partial_{\mu}\partial^{\nu}\right]\int{d}^{4}k\hspace{0.5mm}\Theta(k_{0})\delta(k_{\alpha}k^{\alpha}-m^{2}){\mathcal{A}}_{\nu}e^{-{i}k_{\alpha}x^{\alpha}}+\mathrm{c.c.},
\end{equation}
where the Dirac $\delta$-function $\delta(k_{\mu}k^{\mu})$ and the Heaviside step function $\Theta(k_{0})$ ensure that the dispersion relation $k_{0}=\vert\mathbf{k}\vert$ is satisfied, and ${\mathcal{A}}_{\mu}$ is a vector amplitude given by
\begin{equation}
    {\mathcal{A}}_{\mu}={\mathcal{A}}^{\mathrm{free}}_{\mu}+4\pi{i}e\int{d}\tau\hspace{0.5mm}\dot{u}_{\mu}\Theta(x_{0}-u_{0})e^{ik_{\alpha}u^{\alpha}},
\end{equation}
with $\mathcal{A}_{\mu}^{\mathrm{free}}$ a vector amplitude depending solely on $k_{\mu}$. Having expressed the solution of the field equation in this manner, one can easily observe that the presence of the particle results for the field in a superposition of plane waves with constant amplitudes due to its free propagation and of waves with variable amplitudes whose origin can be shown to be the radiation reaction of the particle (for a detailed proof, see Barut \cite{Barut1980}).

We aim to investigate the possibility of generalizing Cetto's approach to quantization for the electromagnetic radiation field; hence, from this point on, we shall limit ourselves to working in the Lorentz gauge, i.e., taking $\zeta=1$. Then, recalling that the description of the field provided by the MHF in the momentum representation has been shown to be equivalent to its Lagrangian counterpart \cite{JFPB2025}, we have that Eq. (\ref{Hamil1}) can be rewritten as
\begin{equation}
    \mathcal{H}=\frac{1}{8\pi^{3}}\int{d}^{4}k\hspace{0.5mm}\Theta(k_{0})\delta(k_{\alpha}k^{\alpha})\Tilde{\mathcal{H}},\label{HamilMomSpace}
\end{equation}
where
\begin{equation}
    \Tilde{\mathcal{H}}=-\frac{k_{\mu}k^{\mu}}{4\pi{c}k_{0}}{\mathcal{A}}_{\nu}^{\ast}{\mathcal{A}}^{\nu}+\delta_{k}\int{d}\tau\left[p_{\mu}\dot{u}^{\mu}+\frac{m_{0}c^{2}}{2}\right]\delta^{(4)}(x_{\alpha}-u_{\alpha})+\frac{e}{c}\int{d}\tau\left[\mathcal{A}_{\nu}\dot{u}^{\nu}e^{-ik_{\mu}u^{\mu}}+\mathrm{c.c.}\right]\delta(x_{0}-u_{0}),\label{ImDeDonderWeylF}
\end{equation}
with $\delta_{k}$ a $\delta$-function such that $\int{d}^{4}k\hspace{0.5mm}\Theta(k_{0})\delta(k_{\alpha}k^{\alpha})\delta_{k}=8\pi^{3}$, is the image of the system's de Donder–Weyl function in the field's momentum representation. This last function has the same information about the system as $\mathcal{H}$ and therefore constitutes the basis for this alternative description. Furthermore, the structure of Eq. (\ref{ImDeDonderWeylF}) reveals that the de Donder–Weyl function is a Lorentz scalar proportional to the rest mass of the complete system, i.e., particle$+$field\footnote{The origin of the proposed interpretation becomes evident when recalling that the dispersion relation for a massive field is $k_{\mu}k^{\mu}=m^{2}$, with $m$ being the mass parameter associated with the field. For purposes of comparison, examples of de Donder–Weyl functions in the momentum representation for other fields may be seen in Ref. \cite{JFPB2025}}. Naturally, this identification implies a null value for the part of the de Donder–Weyl function that is associated with the free propagation of the electromagnetic radiation field since the dispersion relation states in such a case that $k_{\mu}k^{\mu}=k_{0}^{2}-\vert\mathbf{k}\vert^{2}=0$.

The image of the de Donder–Weyl function in the momentum representation can equivalently be expressed in terms of the field's canonical variables ${q}_{\mu}$ and $\pi_{\mu\nu}$ \cite{JFPB2025}, defined as
\begin{equation}
    {q}_{\mu}=\frac{i(\mathcal{A}_{\mu}-\mathcal{A}_{\mu}^{\ast})}{\sqrt{8\pi{c}k_{0}}},\quad\pi_{\mu\nu}=\frac{k_{\mu}(\mathcal{A}_{\nu}+\mathcal{A}_{\nu}^{\ast})}{\sqrt{8\pi{c}k_{0}}},\label{VarCan}
\end{equation}
and whose Lorentz-covariant Hamilton's equations are
\begin{equation}
    \partial_{\mu}{q}_{\nu}=\frac{\partial\tilde{\mathcal{H}}}{\partial\pi^{\mu\nu}},\quad\partial_{\mu}\pi^{\mu\nu}=-\frac{\partial\tilde{\mathcal{H}}}{\partial{q}_{\nu}},\label{EcsHamiltonQPi}
\end{equation}
respectively. As a consequence, the set of equations constituted by Eqs. (\ref{EcsHamiltonxpparticle}) and (\ref{EcsHamiltonQPi}) completely governs the electrodynamics of the whole system in the Lorentz gauge.

\section{Proposal for a Lorentz-covariant Poisson Bracket}\label{Sec3}

In the Hamiltonian formulation of classical mechanics, the evolution equation of any phase-space function is stated in terms of its PB with the Hamiltonian function of the mechanical system concerned, with Hamilton's equations being special cases of such an evolution equation for the canonical positions and momenta. In particular, when considering the Hamiltonian function itself, one immediately finds that its evolution equation reduces to an equality between its total and explicit time derivatives. So, if we claim that the above description of the system is canonical, we must verify whether the de Donder–Weyl function satisfies a similar mathematical relationship between its derivatives. To this end, we observe that it can be established straightforwardly from Eqs. (\ref{HamilMomSpace}) and (\ref{EcsHamiltonQPi}) that
\begin{equation}
    \partial_{\mu}\mathcal{H}=\partial_{\mu}\mathcal{H}\hspace{0.5mm}\Big|_{\mathrm{explicit}}+\frac{1}{8\pi^{3}}\int{d}^{4}k\hspace{0.5mm}\Theta(k_{0})\delta(k_{\alpha}k^{\alpha})\left[\frac{\partial{\tilde{\mathcal{H}}}}{\partial{q}^{\nu}}\frac{\partial{\tilde{\mathcal{H}}}}{\partial\pi_{\mu\nu}}-\frac{\partial{\tilde{\mathcal{H}}}}{\partial{q}^{\nu}}\frac{\partial{\tilde{\mathcal{H}}}}{\partial\pi_{\mu\nu}}\right],\label{HamilConstant}
\end{equation}
so that we conclude that the de Donder–Weyl function indeed satisfies an equality such as the one mentioned.

The former statement brings PB back to the center of our discussion. As mentioned in Sect. \ref{Sec1}, the MHF accepts multiple ways of defining such geometric objects, yet the structure of Eq. (\ref{HamilConstant}) strongly suggests proposing the bilinear form
\begin{equation}
    \{A,B\}=\int{d}^{4}k\hspace{0.5mm}V_{\mu}\left[\frac{\partial{A}}{\partial{q}^{\nu}}\frac{\partial{B}}{\partial\pi_{\mu\nu}}-\frac{\partial{B}}{\partial{q}^{\nu}}\frac{\partial{A}}{\partial\pi_{\mu\nu}}\right],\label{PoissonBracket}
\end{equation}
where $V_{\mu}$ is to be determined, as the equal-time PB of functions $A$ and $B$ with respect to the canonical variables of the field\footnote{A preliminary version of this proposal can be found in Ref. \cite{JFPB2025}.}. The introduction of $V_{\mu}$ is strictly necessary for $\{A,B\}$ to be antisymmetric and satisfy Leibniz's rule and Jacobi's identity exactly as a PB is required to do (see Refs. \cite{JM1986,MM1998,MM2001,VG2006,JS2008} for proposals with a structure similar to the one presented here). That being the case, applying Eq. (\ref{PoissonBracket}) implies that the equal-time PB of $q_{\mu}$ and $\pi_{\mu\nu}$ is given by
\begin{equation}
    \{q_{\mu}(x_{0},k_{\alpha}),\pi_{\lambda\nu}(x_{0},k_{\alpha}')\}=V_{\lambda}\eta_{\mu\nu}\delta^{(4)}(k_{\alpha}-k_{\alpha}'),\label{ParPoisson}
\end{equation}
from which two conclusions can be drawn. First, and as one could expect, the canonical variables associated with different modes of oscillation of the field are independent of each other. Second, we can rewrite Eq. (\ref{ParPoisson}) so that it resembles the canonical PB of non-relativistic classical mechanics by choosing $V_{\mu}$ in a convenient way.

In order to get a hint on the structure of $V_{\mu}$, it is very enlightening to calculate the equal-time PB of the field and its (generalized) conjugate momentum,
\begin{equation}
    \{A_{\mu}(x_{0},\mathbf{x}),\theta_{\lambda\nu}(x_{0},\mathbf{x}')\}=\frac{1}{2}\left[\frac{\delta(0)}{8\pi^{3}}\right]\eta_{\mu\nu}K_{\lambda}(\mathbf{x},\mathbf{x}')+\mathrm{non}\text{-}\mathrm{free}\hspace{0.5mm}\mathrm{part},\label{PPCampoMomento}
\end{equation}
where we have defined $K_{\mu}$ as
\begin{equation}
    K_{\mu}(\mathbf{x},\mathbf{x}')=\frac{1}{4\pi^{3}}\int{d}^{4}k\hspace{0.5mm}\Theta(k_{0})\delta(k_{\alpha}k^{\alpha})\left[V_{\mu}\sin(\mathbf{k}\cdot\mathbf{x})\sin(\mathbf{k}\cdot\mathbf{x}')+\frac{k_{\alpha}V^{\alpha}}{k_{\alpha}k^{\alpha}}k_{\mu}\cos(\mathbf{k}\cdot\mathbf{x})\cos(\mathbf{k}\cdot\mathbf{x}')\right],
\end{equation}
since this result is independent of the auxiliary variables used to describe the system. Thus, Eq. (\ref{PPCampoMomento}) calls for taking $V_{\mu}$ equal to $ak_{\mu}$, with $a$ a proportionality constant, so that
\begin{equation}
    \{A_{\mu}(x_{0},\mathbf{x}),\theta_{\lambda\nu}(x_{0},\mathbf{x}')\}=\frac{a}{2}\left[\frac{\delta(0)}{8\pi^{3}}\right]\eta_{0\lambda}\eta_{\mu\nu}\delta^{(3)}(\mathbf{x}-\mathbf{x}')+\mathrm{non}\text{-}\mathrm{free}\hspace{0.5mm}\mathrm{part}.\label{PPCampoMomento2}
\end{equation}
With this selection, we easily identify the well-known result from standard field theory for the mentioned PB in the part of the latter equation that is associated with the free propagation of the field. 

The generalization of the PB in Eq. (\ref{PoissonBracket}) to the case of different times is immediate. Following Nikoli\'{c}'s proposal \cite{HN2024}, we establish the non-equal-time PB of functions $A$ and $B$ with respect to the canonical variables of the field as
\begin{equation}
    \{A,B\}({s})=a\int{d}^{4}k\hspace{0.5mm}k_{\mu}\left(\frac{\partial{A}_{s}}{\partial{q}^{\nu}(s)}\frac{\partial{B}_{s}}{\partial\pi_{\mu\nu}(s)}-\frac{\partial{B}_{s}}{\partial{q}^{\nu}(s)}\frac{\partial{A}_{s}}{\partial\pi_{\mu\nu}(s)}\right),\label{PoissonBracketDiffertTime}
\end{equation}
where $A_{s}$ and $B_{s}$ denote that functions $A$ and $B$ have been written in terms of the initial conditions $A(s)$ and $B(s)$, respectively, with $s$ some fixed instant. In the present case, and for the free field, we have that $A_{\mu}$ is a linear function of the initial conditions; therefore, it is straightforward to calculate the non-equal-time PB for $A_{\mu}$ at two different space-time positions, so that we obtain that
\begin{equation}
    \{A_{\mu}(x_{a}),A_{\nu}(x_{a}')\}({s})=2\pi{a}{c}\left[\frac{\delta(0)}{8\pi^{3}}\right]\eta_{\mu\nu}\Delta(x_{\alpha}-x_{\alpha}')+\mathrm{non}\text{-}\mathrm{free}\hspace{0.5mm}\mathrm{part},\label{PPCampoCampoDifTimes}
\end{equation}
where $\Delta$ denotes the Lorentz-invariant Pauli–Jordan function. Similarly, one can directly calculate the non-equal-time PB of the vector potential and its conjugate momentum,
\begin{equation}
    \{A_{\mu}(x_{a}),\theta_{\lambda\nu}(x_{a}')\}({s})=\frac{a}{2}\left[\frac{\delta(0)}{8\pi^{3}}\right]\eta_{\mu\nu}\partial_{\lambda}\Delta(x_{\alpha}-x_{\alpha}')+\mathrm{non}\text{-}\mathrm{free}\hspace{0.5mm}\mathrm{part}.\label{PPCampoMomentoDifTimes}
\end{equation}
Both results are independent of $s$ and coincide, except for a numerical factor\footnote{The factor $\delta(0)/8\pi^{3}$ in Eqs. (\ref{PPCampoMomento}), (\ref{PPCampoMomento2}), (\ref{PPCampoCampoDifTimes}), and (\ref{PPCampoMomentoDifTimes}), or similar factors, will appear when calculating PB with respect to the canonical variables of quantities consisting of sums (integrals) of the contributions of all the modes of oscillation of the field, and is due to the particular choice of the normalization factor for these quantities. This factor could be ``eliminated'' through a renormalization process, but instead of sweeping it under the rug, we have decided to leave it explicit.}, with their counterparts from the conventional formulations of field theories (see, for example, Heitler \cite{WH1954}).


\section{Reduction of the Proposal to Its Non-covariant Form}\label{Sec4}

With the aim of finding an appropriate value for the proportionality constant $a$, let us explore the reduction process of the PB in Eq. (\ref{PoissonBracket}) to its non-covariant form. First, and for purposes that will be clarified later, we rewrite Eq. (\ref{PoissonBracket}) in terms of the amplitudes $\mathcal{A}_{\mu}$ and their complex conjugates, so that we straightforwardly establish that
\begin{equation}
    \{A,B\}_{q\pi}=-4\pi{i}c\{A,B\}_{\mathcal{AA}^{\ast}},
\end{equation}
where we have introduced the subscript $q\pi$ or $\mathcal{AA}^{\ast}$ to denote the set of variables with which the PB is computed and defined $\{A,B\}_{\mathcal{AA}^{\ast}}$ as 
\begin{equation}
    \{A,B\}_{\mathcal{AA}^{\ast}}=a\int{d}^{4}k\hspace{0.5mm}k_{0}\left(\frac{\partial{A}}{\partial\mathcal{A}^{\mu}}\frac{\partial{B}}{\partial\mathcal{A}^{\ast}_{\mu}}-\frac{\partial{B}}{\partial\mathcal{A}^{\mu}}\frac{\partial{A}}{\partial\mathcal{A}^{\ast}_{\mu}}\right).\label{PoissonBracketAA}
\end{equation}
Naturally, Eq. (\ref{PoissonBracketAA}) does not correspond to the standard (non-covariant) PB associated with the normal variables of an electromagnetic radiation field, since for each oscillation mode of the field there is a sum of four terms instead of two.

Nevertheless, we can proceed by assuming that there exists an inertial frame of reference such that we can write each amplitude $\mathcal{A}_{\mu}$ as
\begin{equation}
    \mathcal{A}_{\mu}=\mathcal{A}_{0}\epsilon_{\mu}^{0}+\mathcal{A}_{1}\epsilon_{\mu}^{1}+\mathcal{A}_{2}\epsilon_{\mu}^{2}+\mathcal{A}_{3}\epsilon_{\mu}^{3},\label{GBforA}
\end{equation}
i.e., as a sum of two transversal components, one longitudinal, and one scalar or temporal. In Eq. (\ref{GBforA}), vectors $\epsilon_{\mu}^{i}$, with $i=1,2,3$, are of the form $(0,\mathbf{e}_{i})$ where $\mathbf{e}_{1}$ and $\mathbf{e}_{2}$ denote polarization unit three-dimensional vectors that are orthogonal to each other and orthogonal to $\mathbf{e}_{3}$, a unit three-dimensional vector that is collinear to the spatial component of the wave vector of the mode under consideration. On the other hand, $\epsilon_{\mu}^{0}$ denotes a unit vector of the form $(1,\mathbf{0})$. Hence, we have that Eq. (\ref{PoissonBracketAA}) can be written as
\begin{equation}
    \{A,B\}_{\mathcal{AA}^{\ast}}=a\int{d}^{4}k\hspace{0.5mm}k_{0}\left(\frac{\partial{A}}{\partial\mathcal{A}_{0}}\frac{\partial{B}}{\partial\mathcal{A}^{\ast}_{0}}-\frac{\partial{B}}{\partial\mathcal{A}_{0}}\frac{\partial{A}}{\partial\mathcal{A}^{\ast}_{0}}-\sum_{i=1}^{3}\left[\frac{\partial{A}}{\partial\mathcal{A}_{i}}\frac{\partial{B}}{\partial\mathcal{A}^{\ast}_{i}}-\frac{\partial{B}}{\partial\mathcal{A}_{i}}\frac{\partial{A}}{\partial\mathcal{A}^{\ast}_{i}}\right]\right).\label{PoissonBracketNonRel2}
\end{equation}
Then, since we are working in the Lorentz gauge, we have that the Lorentz auxiliary condition, $\partial_{\mu}A^{\mu}=0$, implies that
\begin{equation}
    \mathcal{A}_{0}-\mathcal{A}_{3}=0,
\end{equation}
so the contributions of the scalar modes to the PB are compensated by those of the longitudinal modes for each oscillation mode of the field. Therefore, we find through this reduction process (Gupta–Bleuler process \cite{SG1950,KB1950}) that the PB in Eq. (\ref{PoissonBracketNonRel2}) is equal to
\begin{equation}
    \{A,B\}_{\mathcal{AA}^{\ast}}=-a\sum_{\lambda=1}^{2}\int{d}^{4}k\hspace{0.5mm}k_{0}\left(\frac{\partial{A}}{\partial\mathcal{A}_{\lambda}}\frac{\partial{B}}{\partial\mathcal{A}^{\ast}_{\lambda}}-\frac{\partial{B}}{\partial\mathcal{A}_{\lambda}}\frac{\partial{A}}{\partial\mathcal{A}^{\ast}_{\lambda}}\right).\label{PoissonBracketNonRel4}
\end{equation}

Now, the PB in Eq. (\ref{PoissonBracketNonRel4}) is certainly not the standard PB for an electromagnetic radiation field, but both expressions are in fact related. To show this, we need to remember that the amplitudes in Eq. (\ref{PoissonBracketNonRel4}) are still four-dimensional and that the relation of these amplitudes with their three-dimensional counterparts is precisely\footnote{See, for example, Heitler \cite{WH1954} or Bogoliubov and Shirkov \cite{NB1959}.}
\begin{equation}
    \mathcal{A}_{\mu}(\mathbf{k})=\frac{\mathcal{A}_{\mu}(k_{\alpha})}{\sqrt{2k_{0}}},\label{RelAmp4d3d}
\end{equation}
so that
\begin{equation}
    \{A,B\}_{\mathcal{AA}^{\ast}}=-\frac{a}{2}\sum_{\lambda=1}^{2}\int{d}^{4}k\left(\frac{\partial{A}}{\partial\mathcal{A}_{\lambda}(\mathbf{k})}\frac{\partial{B}}{\partial\mathcal{A}^{\ast}_{\lambda}(\mathbf{k})}-\frac{\partial{B}}{\partial\mathcal{A}_{\lambda}(\mathbf{k})}\frac{\partial{A}}{\partial\mathcal{A}^{\ast}_{\lambda}(\mathbf{k})}\right).\label{PoissonBracketNonRel5}
\end{equation}
From this result, we immediately establish that
\begin{equation}
    \{A,B\}_{\mathcal{AA}^{\ast}}=-\frac{a}{2}\int{d}k_{0}\left[\sum_{\lambda=1}^{2}\int{d}^{3}k\hspace{0.5mm}\left(\frac{\partial{A}}{\partial\mathcal{A}_{\lambda}(\mathbf{k})}\frac{\partial{B}}{\partial\mathcal{A}^{\ast}_{\lambda}(\mathbf{k})}-\frac{\partial{B}}{\partial\mathcal{A}_{\lambda}(\mathbf{k})}\frac{\partial{A}}{\partial\mathcal{A}^{\ast}_{\lambda}(\mathbf{k})}\right)\right].\label{PoissonBracketNonRel6}
\end{equation}
Thus, we find that our proposal is equal to the integral over $k_0$ of the standard PB for an electromagnetic radiation field if we take $a=-2$ and identify $\lambda$ as the polarization index. A relationship involving an integral is to be expected since the standard PB involves an integral over the field's three-dimensional wave vector, which is not Lorentz-covariant, so this extra operation guarantees the covariance of our proposal.

It is very important to note that although our proposal indeed contains the standard PB for an electromagnetic radiation field, we have, due to Eq. (\ref{RelAmp4d3d}), that
\begin{equation}
    \{\mathcal{A}_{\mu}(k_{\alpha}),\mathcal{A}^{\ast}_{\nu}(k_{\alpha}')\}_{\mathcal{AA}^{\ast}}\neq\{\mathcal{A}_{\mu}(\mathbf{k}),\mathcal{A}^{\ast}_{\nu}(\mathbf{k}')\}_{\mathcal{AA}^{\ast}},\label{DefDefPP}
\end{equation}
with 
\begin{equation}
    \{\mathcal{A}_{\mu}(k_{\alpha}),\mathcal{A}^{\ast}_{\nu}(k_{\alpha}')\}_{\mathcal{AA}^{\ast}}=-2k_{0}\eta_{\mu\nu}\delta^{(4)}(k_{\alpha}-k_{\alpha}').\label{ParPoissonAA}
\end{equation}
The PB in the right-hand side of Eq. (\ref{DefDefPP}) is generally used as the starting point for canonical quantization in the conventional formulations of quantum electrodynamics. This primacy is the reason why the reduction process of the proposed PB was done in terms of amplitudes rather than canonical variables. It is particularly interesting to note that our result for the PB of two amplitudes of the electromagnetic radiation field, Eq. (\ref{ParPoissonAA}), coincides in structure with that obtained by Dirac through a completely different approach \cite{PD1949}.

\section{Joint Description of the Electrodynamics of a Point Particle}\label{Sec5}

As stated in Ref. \cite{JFPB2025}, given the resemblance of the result of Eq. (\ref{ParPoisson}) with the canonical PB of classical mechanics, one can envision the possibility of defining a single PB associated with the canonical variables of the complete system. In order to construct such an object, let us first examine how to construct a Lorentz-covariant PB with respect to the canonical variables of the particle. One can find in the literature that this problem has been addressed on multiple occasions for purposes similar to those explored here (for example, see Ref. \cite{PD1949,IT1976,FR1979} and references therein). In any case, it is straightforward to propose the bilinear form
\begin{equation}
    \{A,B\}_{up}=-\left(\frac{\partial{A}}{\partial{u}_{\mu}}\frac{\partial{B}}{\partial{p}^{\mu}}-\frac{\partial{B}}{\partial{u}_{\mu}}\frac{\partial{A}}{\partial{p}^{\mu}}\right),\label{PoissonBracketPart}
\end{equation}
as the equal-time PB of $A$ and $B$ with respect to the canonical variables of the particle (the generalization to the case of different times is direct from Nikoli\'{c}'s proposal). Consequently, we have that the equal-time PB of the canonical variables of the mechanical subsystem is given by
\begin{equation}
    \{u_{\mu},p_{\nu}\}_{up}=-\eta_{\mu\nu}.\label{PoissonBracketPartCan}
\end{equation}

On this basis, we establish that the set $(Q;P)$ of canonical variables of the point particle+electromagnetic radiation field system should be the union of the sets of canonical variables of the mechanical subsystem and the field, that is,
\begin{equation}
    (Q;P)=(u_{\mu},q_{\mu}(k_{\alpha});p_{\mu},\pi_{\mu\nu}(k_{\alpha})).
\end{equation}
Hence, we define
\begin{equation}
    \{A,B\}_{QP}(s)=\{A,B\}_{up}(s)+\{A,B\}_{q\pi}(s)
\end{equation}
as the PB associated with all the canonical variables at instant $s$ and postulate such bracket as the basis for a joint description of the complete system that is both canonical and Lorentz-covariant.

An interesting feature of the proposed description comes from the fact that the canonical variables of the system at any two different instants, namely $s$ and $\tau>s$, are related through a canonical transformation. Thus, we have that
\begin{equation}
    \{u_{\mu}(\tau),p_{\nu}(\tau)\}_{QP}(s)=\{u_{\mu}(\tau),p_{\nu}(\tau)\}_{QP}(\tau),\label{TransCan}
\end{equation}
since PB are invariant under those transformations. Therefore, we straightforwardly establish from Eqs. (\ref{PoissonBracketPartCan}) and (\ref{TransCan}) that
\begin{equation}
    \{u_{\mu}(\tau),p_{\nu}(\tau)\}_{up}(s)+\{u_{\mu}(\tau),p_{\nu}(\tau)\}_{q\pi}(s)=-\eta_{\mu\nu},\label{TransCan2}
\end{equation}
i.e, that the dependence of the canonical variables of the mechanical subsystem at instant $\tau$ (or at any other instant) on the initial conditions $u_{\mu}(s)$, $p_{\mu}(s)$, $q_{\mu}(s)$, and $\pi_{\mu\nu}(s)$ is precisely of the form necessary to ensure that their PB equals the metric tensor.

\section{Conclusions}

The apparent incompatibility between the requirement of field theories for locally describing the interactions between particles and the extended use of the canonical quantization process seems to have found, at least in principle, a solution in the MHF. As we showed in this article, it is feasible to use this formalism to construct a description of the complete (particle$+$electromagnetic radiation field) system that is suitable for a canonical treatment and, at the same time, treats all space-time coordinates symmetrically. The pivotal element of such a description is an unambiguous Lorentz-covariant PB, Eqs. (\ref{PoissonBracket}) and (\ref{PoissonBracketDiffertTime}), which not only reproduces the well-known results from the conventional formulations of field theories, but also contains the standard (non-covariant) PB for an electromagnetic radiation field when reduced to its non-covariant form. On this basis, we obtained the PB of the canonical variables of the particle and of the field, Eqs. (\ref{ParPoisson}) and (\ref{PoissonBracketPartCan}). These results constitute a generalization of the canonical PB of non-relativistic classical mechanics and might be taken as the departure point to generalize Cetto's approach to the quantization of the electromagnetic radiation field and, therefore, develop a manifestly Lorentz-covariant quantization process for such a field.

The generalization of the PB in Eqs. (\ref{PoissonBracket}) and (\ref{PoissonBracketDiffertTime}) for arbitrary values of the gauge-fixing parameter, other than that associated with the Lorentz gauge, remains to be done in order to construct a Lorentz-covariant generalization of canonical quantization for general fields.
\vspace{0.5cm}

\noindent \textbf{Acknowledgments:} The author would like to thank two anonymous reviewers for their very helpful comments.
\vspace{0.5mm}

\noindent \textbf{Funding information:} Not applicable.

\end{document}